\documentclass[twocolumn,twoside,slac]{revtex4}

\usepackage{graphicx}

\usepackage{fancyhdr}

\usepackage[footnote]{acronym}

\begin{document}


\title{An on-line Integrated Bookkeeping: electronic run log book and
  Meta-Data Repository for ATLAS}


%




\author{M. Barczyc}
\affiliation{CERN, Geneva, Switzerland}
\author{D. Burckhart-Chromek}
\affiliation{CERN, Geneva, Switzerland}
\author{M. Caprini}
\affiliation{CERN, Geneva, Switzerland}
\author{J. Da Silva Conceicao}
\affiliation{CERN, Geneva, Switzerland}
\author{M. Dobson}
\affiliation{CERN, Geneva, Switzerland}
\author{J. Flammer}
\affiliation{CERN, Geneva, Switzerland}
\author{R. Jones}
\affiliation{CERN, Geneva, Switzerland}
\author{A. Kazarov}
\affiliation{CERN, Geneva, Switzerland}
\author{S. Kolos}
\affiliation{CERN, Geneva, Switzerland}
\author{D. Liko}
\affiliation{CERN, Geneva, Switzerland}
\author{L. Mapelli}
\affiliation{CERN, Geneva, Switzerland}
\author{I.Soloviev}
\affiliation{CERN, Geneva, Switzerland}

\author{R. Hart}
\affiliation{NIKHEF, Amsterdam, Netherlands}

\author{A. Amorim}
\affiliation{CFNUL/FCUL, Universidade de Lisboa, Portugal}
\author{D. Klose}
\affiliation{CFNUL/FCUL, Universidade de Lisboa, Portugal}
\author{J. Lima}
\affiliation{CFNUL/FCUL, Universidade de Lisboa, Portugal}
\author{L. Lucio}
\affiliation{CFNUL/FCUL, Universidade de Lisboa, Portugal}
\author{L. Pedro}
\affiliation{CFNUL/FCUL, Universidade de Lisboa, Portugal}
\author{H. Wolters}
\affiliation{UCP Figueira da Foz, Portugal}

\author{E. Badescu}
\affiliation{NIPNE, Bucharest, Romania}

\author{I. Alexandrov}
\affiliation{JINR, Dubna, Russian Federation}
\author{V. Kotov}
\affiliation{JINR, Dubna, Russian Federation}
\author{M. Mineev}
\affiliation{JINR, Dubna, Russian Federation}

\author{Yu. Ryabov}
\affiliation{PNPI, Gatchina, Russian Federation}

\begin{abstract}
In the context of the ATLAS experiment there is growing evidence of the importance
of different kinds of Meta-data including all the important details of the
detector and data acquisition that are vital for the analysis of the acquired
data. The Online BookKeeper (OBK) is a component of ATLAS online software
that stores all information collected while running the experiment, including
the Meta-data associated with the event acquisition, triggering and
storage. The facilities for acquisition of control data within the on-line
software framework, together with a full functional Web interface, make the
OBK a powerful tool containing all information needed for event analysis,
including an electronic log book.

In this paper we explain how OBK plays a role as one of the main collectors
and managers of Meta-data produced on-line, and we'll also focus on the Web
facilities already available. The usage of the web interface as an electronic
run logbook is also explained, together with the future extensions.

We describe the technology used in OBK development and how we arrived at the
present level explaining the previous experience with various DBMS
technologies. The extensive performance evaluations that have been performed
and the usage in the production environment of the ATLAS test beams are also
analysed.  

\end{abstract}
\maketitle
\thispagestyle{fancy}





\section{\label{Intro}INTRODUCTION}
Experiments in High Energy Physics (HEP) are becoming increasingly more
complex. The construction of a new particle accelerator and associated
detectors is a technological challenge that also encompasses the development of
an associated software system.

The Large Hadron Colider (LHC) currently being built at the European
Organization for Nuclear Research (CERN), will support four different detectors
installed. The ATLAS \cite{atlas} (A Toroidal LHC ApparatuS) at LHC will
require a complex trigger system. This trigger will have to reduce
the original 40MHz of p-p interaction rate to a manageable 100Hz for
storage. The total mass storage, including raw, reconstructed, simulated and
calibration data exceeds 1 PetaByte per year. The three major types of data to
be stored by ATLAS are:
\begin{itemize}
\item{\it{raw event data}}, data collected by the detector resulting from the
  particle collisions (events) 
\item{{\label{conditions}\it{conditions and Meta-data}}}, which includes
  calibration constants, run   conditions, accelerator conditions, trigger
  settings, detector configuration, and Detector Control System (DCS)
  conditions that determine the   conditions under which every physics event
  occurred. These conditions are stored with an interval of validity (typically
  time or run number) and retrieved using time (or run number) as a key
  \cite{paoli}. Also the DAQ status and time evolution of the Configuration
  Database is included in this type of data.
\item{\it{reconstructed data}}, corresponding to objects with
  physical meaning (e.g. electrons, tracks, etc.) that are the result of
  applying software algorithms to the raw data, taking into consideration the
  conditions under which the raw data was taken. 
\end{itemize}

An important part of the conditions database data is associated with the
Trigger and Data Aquisition System \cite{daq} (TDAQ). Through the TDAQ system
flow very different types of data (e.g. calibration and alignment information,
configuration databases information) \cite{cdi} that
appear as conditions data for storage. Some types of this data, such as the 
accelerator's beam parameters, detector configuration, test-beam table
position, are recorded by a specific component of the TDAQ system, the Online
Bookkeeper (OBK). The OBK is important as a source of data for the conditions
database, as an entry point to analysis jobs on the raw data and as a debug
resource for the DAQ system. For this reason, OBK can be qualified as
one of the biggest collectors and managers of Meta-data produced online for
the ATLAS experiment. The aim of the OBK is to store information describing
the data acquired by the DAQ and to provide offline access to this information \cite{obk-highlevel}. It is also a powerful tool for users in the
control room who can use it as a Run log book to attach their comments, or
other types of support information. 

\section{OBK AND THE  ONLINE SOFTWARE}
This section will provide an overview of
how OBK works in the Online Softare framework. As OBK is a software package
of the Online Software system for the ATLAS TDAQ, the architecture of both, Online Software and OBK, will be briefly described.

\subsection{The Online Software architecture}
 The role of the Online Software is to provide to other TDAQ systems,
configuration, control and monotoring services. It does not include the
processing and transportation of physics data. All packages of the
Online Software create a framework generic enough to allow supervision of many
distinct data taking configurations. From the architectural point of view there are
three different \textit{group of components}: \textit{group of components} are the \textbf{Configuration},
\textbf{Control} and \textbf{Monitoring}. Table \ref{tab-online}
shows these main packages and their components.

{\tiny{
\begin{table}[h]

\begin{center}

\caption{Online software packages and its components}

\begin{tabular}{|c|}

\hline \textbf{Configuration} \\

\hline Configuration Databases \\
 Online Bookkeeper \\
\hline
\hline \textbf{Control} \\
\hline Run Control \\
 DAQ Supervisor \\
 Process Manager \\
 Ressource Manager \\
 Integrated Graphical User Interface \\
\hline
\hline \textbf{Monitoring} \\
\hline Information Service \\
 Message Reporting System \\
 Online Histograming \\
 Event Monitoring \\
\hline

\end{tabular}

\label{tab-online}

\end{center}

\end{table}
}
}
\subsection{The OBK architecture}
\begin{center}
\begin{figure}[h]
\centering
\includegraphics[width=7.5cm]{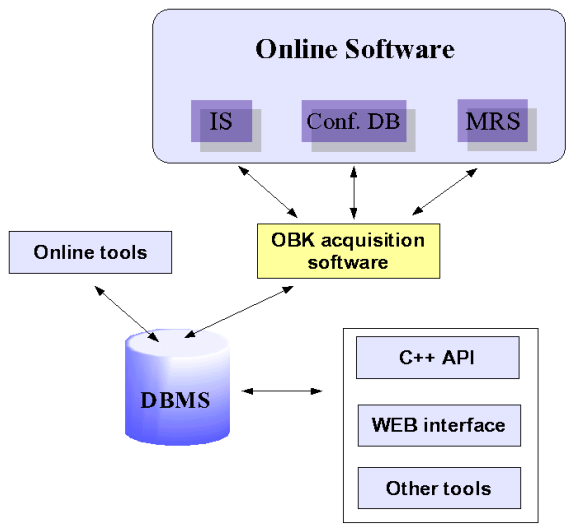}
\caption{Generic OBK architecture.} \label{gen-arch}
\end{figure}
\end{center}

The OBK is part of the \textbf{Configuration} group. A first prototype was
developed 4 years ago and since it has evolved significantly. Figure
\ref{gen-arch} shows a simplified diagram of OBK in which are the other
Online packages that OBK interacts with. Namely, the Information Service (IS),
Message Reporting System (MRS) and Configuration Databases (Conf. DB). 

Both IS and MRS use the IPC (Inter Process Comunication) package,
a CORBA implementation, as messaging backbone \cite{corba}. MRS provides the facility which allows all software
components in the ATLAS DAQ system and related processes to report error
messages to other components of the distributed TDAQ system. IS provides the
possibility for inter-application information exchange in the distributed
environment.
These three components of the Online SW are the providers of information to OBK
databases. The data stored in OBK databases the information will be
automatically available worldwide. This process begins with the data
acquisition process in the distributed environment and ends up with
the final users that will query the database for the most relevant
informations about each data taking period.

Figure \ref{gen-arch} also shows schematically a general architecture of
OBK. The OBK acquisition software subscribes to relevant MRS and IS servers in
order to receive MRS/IS massages exchanged between components via CORBA/IIOP
callbacks. The information is then stored using an per-run basis
philosophy. The information stored includes the date/time of each run, 
the basic physics parameters, the status of a run - if it was a successfully
completed or not, etc.

The data is then made available for offline usage. The users dispose of a set
of tools like the Web interface and the C++ API which can be used to interact with
OBK databases. They can retrieve data for a particular Run, or a set of Runs,
or append new relevant information if needed. The web browser can display all
the information of a Run including all IS informations stored, and with the
C++ API there is also the facility to search for instances of a particular IS
class parameter or to cicle all the run headers (StartDate, EndDate,
TriggerType, etc.) in a partition.

\section{Prototyping with OBK}
During the last 4 years, three different OBK prototypes were implemented. The
aim was both to learn with the experience while trying to implement a package
that meets the requirements \cite{obk-requirements}, and to use different Data
Base Management Systems (DBMS). This provided a better understanding of which
DBMS most fulfills the OBK needs inside the Online Software framework. With this
multi-technology approach we're gaining technological expertise about
different DBMS technologies and are then able to make a solid
recommendation for a technological and design solution for a production
bookkeeper tool. This prototyping approach seems to achieve very good results in a
long term experience like this one because each of the new OBK prototypes
provides more functionality and performance gains over its
predecessor. All of the prototypes use C++ as programming language and the
same basic architecture but different DBMS for the persistent storage. 

\subsection{Objectivity/DB based}
At the begining, LHC experiments selected Objectivity/DB for the official DBMS
persistent storage.  An OBK implementation using this DBMS was then the first natural choice.

This prototype was designed to take full advantage of the pure Object
Oriented (OO) model of the DBMS. The model used with the Objecivity/DB is
organized in federations, databases, containers and object, OBK was able
to map the data coming from the TDAQ system in a repository structured in
such a way. 

The OBK Objectivity/DB based prototype was used in 2000 testbeams with
success. For data retrieval a web browser was also developed. This allows
users to get in `touch` with the data in a very natural fashion. Since 2001,
this prototype was abandoned and is no longer maintained.

\subsection{OKS based}
The second prototype implemented uses OKS as a persistency mechanism. OKS is a
C++ based, in-memory persistent object manager developed as a package inside
the Online Software \cite{oks}. OKS is also OO but not so sophisticated as
Objectivity/DB. The data files used for storage are created as XML
files. They are stored in the file system which can be local, AFS or NFS. The
access to the data is done by reading directly the files and not through a
centralized server. The usage of XML files to store data is a very interesting
feature because they are human readable and also highly portable.

The main reasons that led to this implementation were
the fact that OKS is \textit{Open Source} software (which makes it usable in
any place without having any problems with licencing) and also that this DBMS
is lighter and more oriented to systems with very high demands in terms of
performance than the Objectivity/DB. There are of course some disadvantages of
using OKS, mainly the lack of features that other DBMS provide like transactions. 

The persistent object schema of this prototype is very similar to the first
one beacause it is also OO featured and the intrinsic data storage philosophy
can be very similar. A web browser with the same approach as the one from
Objectivity is also provided. 

This prototype provides extended features while compared to the first one,
such as more programs to control the databases and a full featured C++ API. All
it's new improvements were used in 2001 test beams at CERN by the users. It
behaved well and acquired several Megabytes of data to the local file system
of a machine and later on AFS. Despite the good behavior from OBK the problem
of data dispersion and consistency soon arrived because this prototype uses several XML files to
store information about each Run.

\subsection{MySQL based}
This is the only OBK prototype that uses a Relational DBMS. MySQL \cite{mysql}
is a well
known, fast and reliable \textit{Open Source} DBMS. It started a new phase in
the OBK development - The phase of the relational model. The decision to
implement a package such as OBK using MySQL was driven from the power of its
underlying SQL engine and also due to the desire of trying a relational
approach to OBK databases. Technically a new database schema was implemented to
allow mapping of data coming mainly from OO sources forcing us to
completely redesign the internal structure of OBK. We have achieved a mapping
between an OO and a Relational schema that is suitable for OBK needs and
started to use it.

This prototype provides all the features of the previous ones, plus further enhancements regarding the users needs and also the very important aspect
of performance. In this implementation the concept of log book was also
introduced and successfully deployed (see section \ref{sec-logbook} for details) and
successfully used.

The MySQL implementation was used successfully in the 2002 test beam. It
recorded more than 1 GByte of data, including data coming from the DCS.
\section{OBK INTERFACES FOR DATA RETRIEVAL}
OBK provides a set of interfaces that allow users to interact with the
databases. There are also available some tools coded in C++ which make some
tasks very easy to execute. In this section the focus will be on the C++
interfaces (the Query API) and on the Web interface used to store information
directly related to the users.
\subsection{C++ Query API}
Both OKS and MySQL prototype are distributed with a C++ Query API. This API
exposes methods that allow data retrieval in a very user oriented
approach. The API encapsulates all the necessary mechanisms to get the correct
values from the OBK databases. The users do not need any special skill (like
how to perform SQL queries) regarding the database schema. This also allows to
preserve the integrity of the database because it doesn't allow users to
manipulate it directly in what concerns for example the addiction of new
information retated to a Run. 
  
\subsection{\label{sec-logbook}Web Browser}
The latest version of the OBK browser includes several functionalities making
it a powerful tool. It provides not only the functionalities of displaying the data coming
from the OBK data acquisition process just after it was collected through the
Online System but also the possibility of behaving like a Run Log Book. It was
built with an excellent searching mechanism oriented to the final users which
will be the Physicists. 
\begin{figure}[h]
\centering
\includegraphics[width=7.5cm]{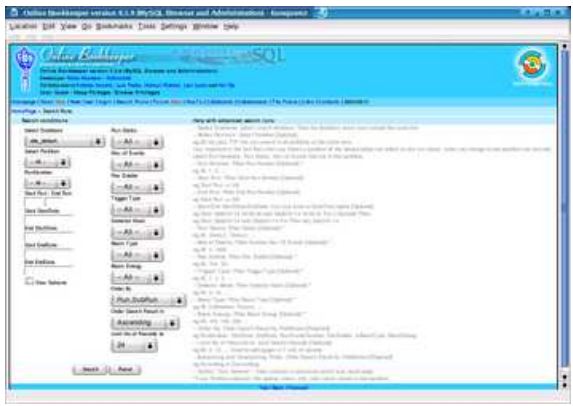}
\caption{OBK browser - search mechanism.} \label{obk-websearch}
\end{figure}\\

\begin{figure}[h]
\centering
\includegraphics[width=7.5cm]{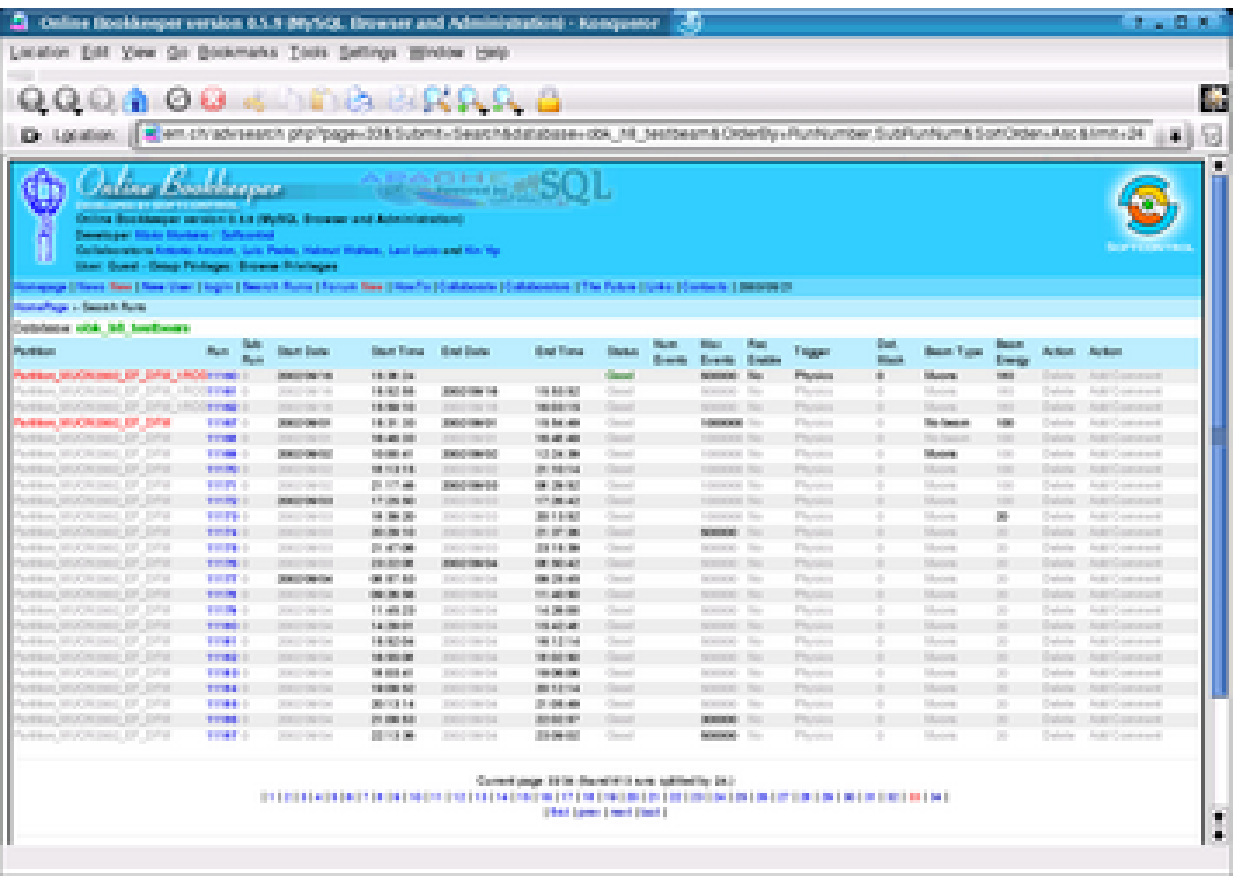}
\caption{OBK browser - result of the search.} \label{obk-seeruns}
\end{figure}

Figure \ref{obk-websearch} shows all the
options of this search mechanism. There are a set of options on
which the users can base their criteria of selection such as:\\
\begin{itemize}
\item \textbf{RunStatus} Good/Bad
\item \textbf{MaxEnvents} The maximum number of events of the Runs
\item \textbf{Start Date/End Date}. For the Start and end of a Run.  This
  allows to `map' the obk run bases data type in a time interval
\item \textbf{BeamType} Muons, Electrons, etc.
\item \textbf{TriggerType} Cosmic, Calibration, Physics
\end{itemize}
Sorting options are also included. The OBK data display in this browser was driven by the need of a clear and very
user friendly interface where the data would be easily accessible to the
users. After the selection of all the criteria that the user wants to meet, 
the result will appear in another web page similar to the one in Figure
\ref{obk-seeruns}. 

The result presents to the users some relevant information
about each Run: Partition to which each Run belongs; Run Number; Start and End
Date of the Run; Run Status; Number of Events; Maximum Number of Events;
Trigger Type, Detector Mask and  Beam Type. It is also possible to display
more specific
information, like for example which messages from the MRS were transfered between
the various components of the Online Software. This is accessible by following
the link provided for each Run. 

\begin{figure}[h]
\centering
\includegraphics[width=7.5cm]{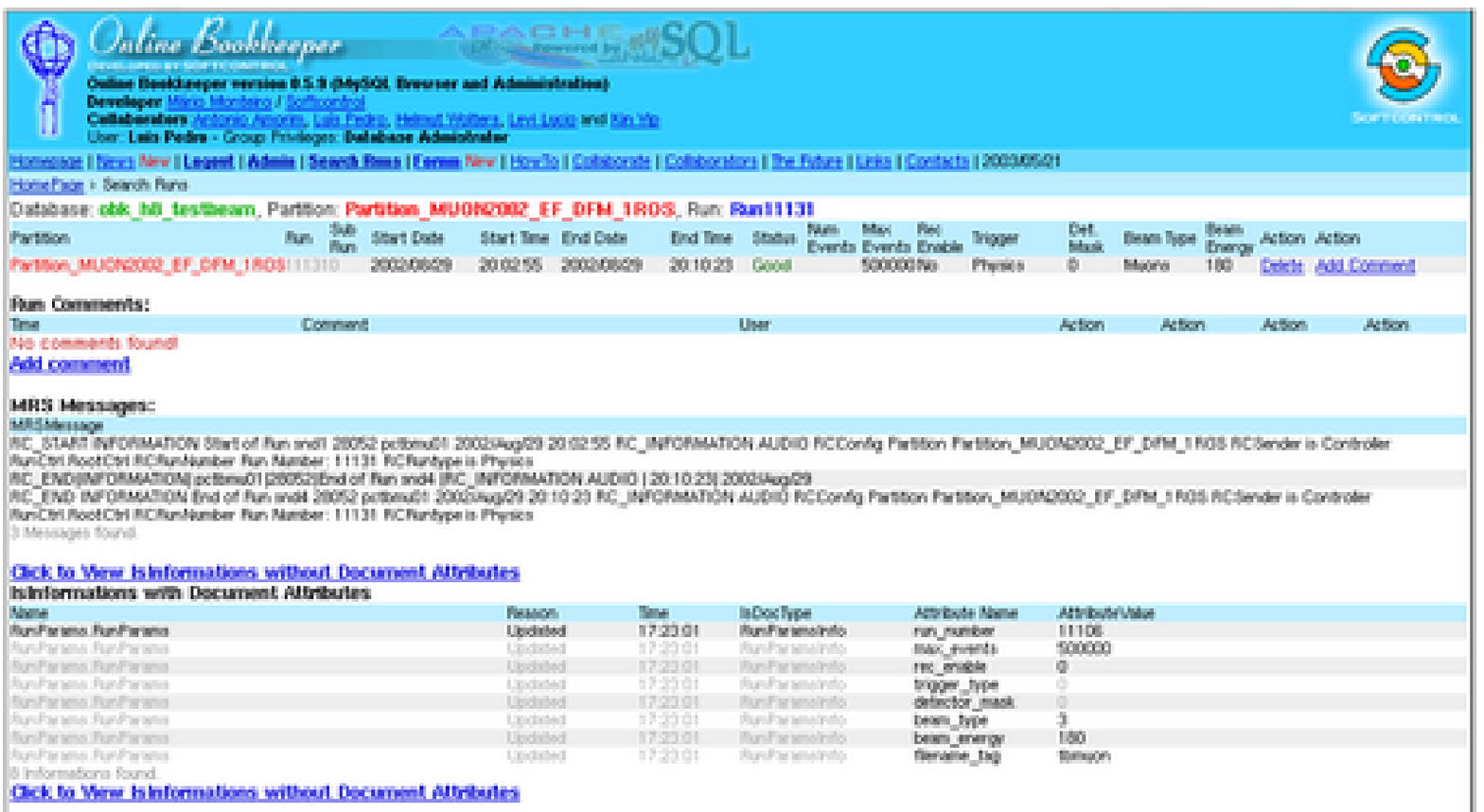}
\caption{OBK browser - more information about a particular Run.} \label{obk-insiderun}
\end{figure}

Figure \ref{obk-insiderun} shows a typical page generated when selecting the
option to display more detailed information on a particular Run. Through this
new page it is possible to browse information including the
messages from the MRS, from the IS and the users comments and attached
files. There are two different approaches to storing comments in OBK
databases: 
\begin{itemize}
\item using binary programs provided for both online and offline comments
\item Using the web browser
\end{itemize}

\begin{figure}[h]
\centering
\includegraphics[width=6cm]{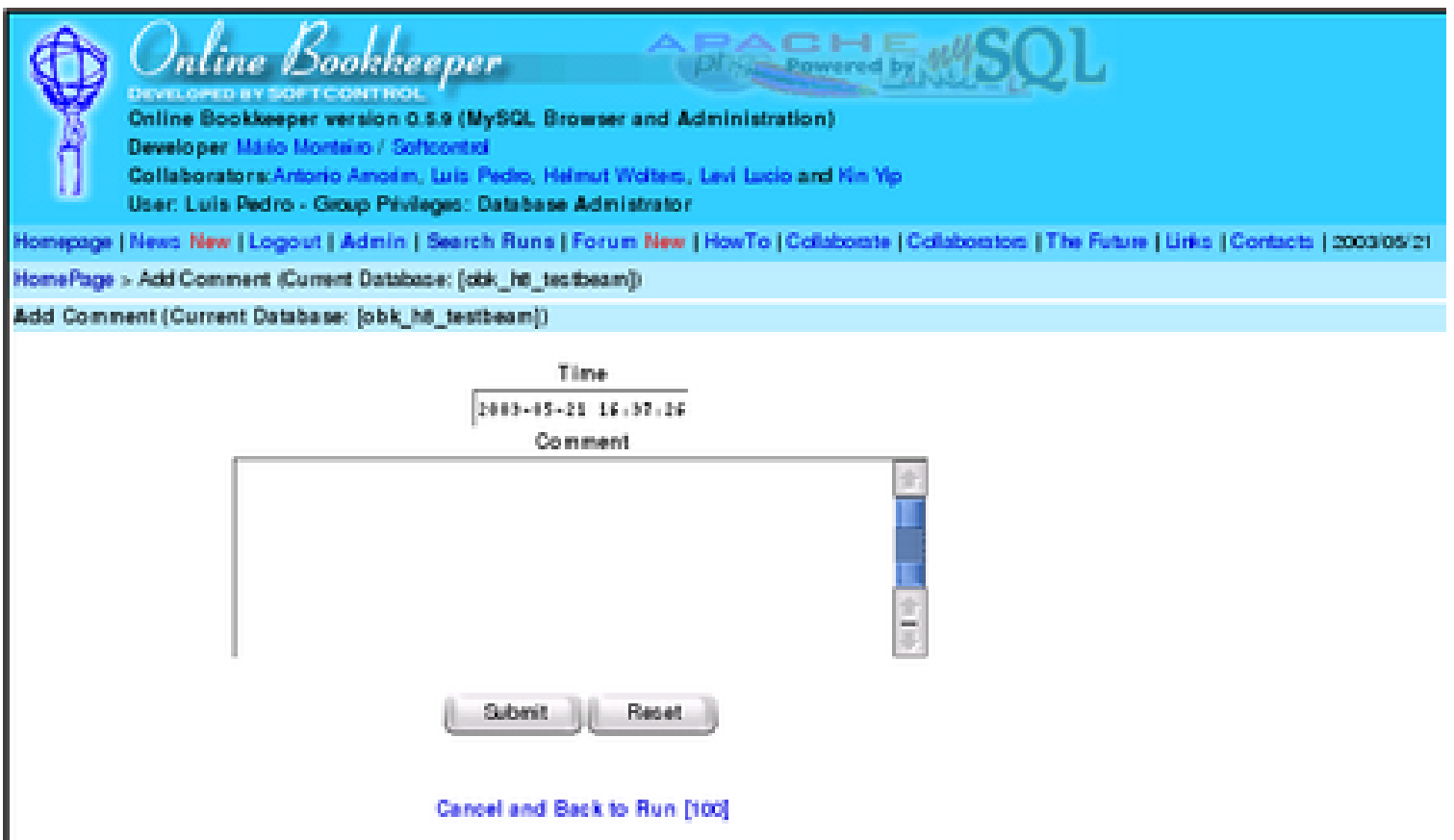}
\caption{OBK browser - adding a comment.} \label{obk-comment}
\end{figure}
Adding comments through the web browser gives also the option of attaching any
type of files to a comment. This allows users to add information that they
think might be relevant to the Run. Afterwards every one can
see the comments and their respectively attached files. Some types of files are
supported and will immediately be displayed in a
different window. File types that are not supported will have to be downloaded
and the appropriate program must be used to open them. 

The OBK browser itself also provides administrative tools, for example to
create databases with the correct structure for OBK, and gives a set of options
for user management: authentication, permissions, etc. 

\section{TESTING}
For the evaluation of the various prototypes the focus was to analyse the
different functionalites, how easy it was to map the data coming from the
Online Sytem or on how complex the code of each one became. On all of
these issues a set of scalablity and performance tests were addressed. One
other objective of these tests was also to evaluate if OBK can handle all the
information produced in the final system. 

\begin{figure}[h]
\centering
\includegraphics[width=7.cm]{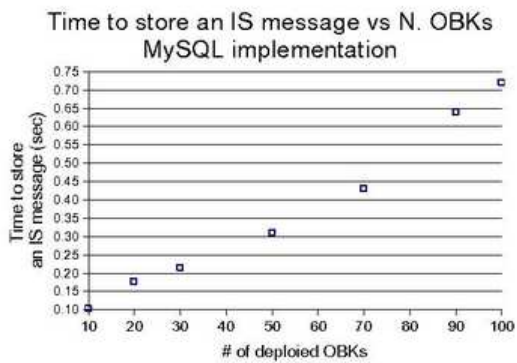}
\caption{OBK tests - scalability test performed to the MySQL implementation.} \label{obk-performance}
\end{figure}

In Figure \ref{obk-performance}  it can be seen the time to store a typical IS
message in function of the number of OBK data acquisition programs running
simultaneously. This test was one of the tests performed in the scalability
context regarding the final system. 

A comparison between the three prototypes was also addressed. This test was
performed with a typical Start of Run message coming from the MRS. The time
for this test is not only the time spent to store the message itself but also
all operations that this implies. This operation includes the creation of a new
Run in the database with all associated operations like the creation of new
files (in the OKS prorotype) or containers (in the case of the Objectivity/DB
prototype). More details about this tests can be found in \cite{obk-chep01}.
The MySQL prototype proved to be the faster one while the Objectivity/DB is
the slowest. It was clear from the test that there is a dependency on the Run
number for the time spent to store a message of this kind in case of both
Objectivity/DB and OKS prototypes. The slope of the OKS line is less than the
Objectivity/DB but when a transaction becomes commited, the Objectivity/DB
prototype gets better. We attribute these results not only to the evolution
in the design from prototype to prototype but also because MySQL provides a
faster engine that makes the time to store these messages negligible
when compared with the other prototypes. Some other performance results are
displayed in tables \ref{tab-perf1} and \ref{tab-perf2}.

{\tiny{
\begin{table}[h]

\begin{center}

\caption{Time (in miliseconds) to Start of Run, End of Run, Comment and
  a typical IS message for \textbf{OBK/OKS} prototype.}

\begin{tabular}{|c|c|c|c|c|}

\hline \textbf{Platform} & \textbf{Start Run} &
\textbf{End Run} & \textbf{Comment} & \textbf{IS}\\

\hline Linux/egcs1.1 & 77-259 & 48-245 & 0,3 & 2,7\\
\hline Linux/gcc2.96 & 47-196 & 29-184 & 0,3 & 1,9 \\

\hline

\end{tabular}

\label{tab-perf1}

\end{center}
\end{table}
}
}

The results presented for the Start of Run, End of Run and Comment, represent
the the minimum value and the maximum observed during the tests. Tests were
performed to a maximum of 500 Runs. For the IS time it's the mean time
to store a IS message from OBK because it was observed that there was no significant growth. 

{\tiny{
\begin{table}[h]

\begin{center}

\caption{Time (in miliseconds) to Start of Run, End of Run, Comment and
  a typical IS message for \textbf{OBK/MySQL} prototype.}

\begin{tabular}{|c|c|c|c|c|}

\hline \textbf{Platform} & \textbf{Start Run} &
\textbf{End Run} & \textbf{Comment} & \textbf{IS}\\

\hline Linux/egcs1.1 & 0,004 & 0,010 & 0,002 & 0,011\\
\hline Linux/gcc2.96 & 0,018 & 0,021 & 0,004 & 0,020 \\

\hline

\end{tabular}

\label{tab-perf2}

\end{center}
\end{table}
}
}
More information about these tests can be found in OBK test
report \cite{obk-perf}. This document includes a detailed description of the
test procedure and other results such as functionality tests.

\section{SUMARY AND FUTURE WORK}
In this paper we presented a general overview of our
experience of using different DBMSs in prototyping
OBK in the ATLAS Online Software framework.
Since the beginning of the project we tried to understand
the problem of bookkeeping for the ATLAS
experiment. For that reason, OBK evolved and it
now provides some tools that can be
qualified as Run log book tools. The last version of
OBK which is the most performant and robust one is
the result of this experience.
But OBK work is still in progress. Included in the list of
future improvements for OBK are:

\begin{itemize}
\item to make it DBMS independent in order to have a more general and dynamic
architecture; 
\item create a set of new tools to extend the actual existing ones for
the log book approach; 
\end{itemize}
\begin{acknowledgments}
The authors wish to thank Mario Monteiro from
\textit{Softcontrol}\cite{soft} who is the main
developer of the OBK MySQL implementation web browser.\\\\\\
\end{acknowledgments}




\end{document}